\documentclass[aps,twocolumn,pra,superscriptaddress,amsmath,showpacs,tightenlines]{revtex4-1}
\usepackage{amssymb}
\usepackage{amsmath}
\usepackage{graphicx}
\usepackage{subfigure}
\usepackage{natbib}
\usepackage{epsfig}
\usepackage{amsfonts}
\usepackage{mathrsfs}
\usepackage{xcolor}
\usepackage[toc,page,title,titletoc,header]{appendix}

\begin{document}

\title{Quantum metrology enhanced by coherence-induced-driving in a cavity QED setup}
\author{Weijun Cheng}
\email{W. Cheng and S. C. Hou contributed equally to this paper.}
\affiliation{Center for Quantum Sciences and School of Physics, Northeast Normal University, Changchun 130024, China}
\author{S. C. Hou}
\email{W. Cheng and S. C. Hou contributed equally to this paper.}
\affiliation{Department of Physics, Dalian Maritime University, Dalian 116026, China}
\affiliation{Center for Quantum Sciences and School of Physics, Northeast Normal University, Changchun 130024, China}
\author{Zhihai Wang}
\email{wangzh761@nenu.edu.cn}
\affiliation{Center for Quantum Sciences and School of Physics, Northeast Normal University, Changchun 130024, China}
\affiliation{Center for Advanced Optoelectronic Functional Materials Research, and Key Laboratory for UV-Emitting Materials and Technology of
Ministry of Education, Northeast Normal University, Changchun 130024, China}
\author{X. X. Yi}
\email{yixx@nenu.edu.cn}
\affiliation{Center for Quantum Sciences and School of Physics, Northeast Normal University, Changchun 130024, China}
\affiliation{Center for Advanced Optoelectronic Functional Materials Research, and Key Laboratory for UV-Emitting Materials and Technology of
Ministry of Education, Northeast Normal University, Changchun 130024, China}

\begin{abstract}
We propose a quantum metrology scheme in a cavity QED setup to achieve the Heisenberg limit. In our scheme, a series of identical two-level atoms randomly pass through and interact with a dissipative single-mode cavity. Different from the entanglement based Heisenberg limit metrology scheme, we do not need to prepare the atomic entangled states before they enter into the cavity.  We show that the initial atomic coherence will induce an effective driving to the cavity field, whose steady state is an incoherent superposition of orthogonal states, with the superposition probabilities being dependent on the atom-cavity coupling strength. By measuring the average photon number of the cavity in the steady state, we demonstrate that the root-mean-square of the fluctuation of the atom-cavity coupling strength is proportional to $1/N_c^2$ ($N_c$ is the effective atom number interacting with the photon in the cavity during its lifetime). It implies that we have achieved the Heisenberg limit in our quantum metrology process. We also discuss the experimental feasibility of our theoretical proposal. Our findings may find potential applications in quantum metrology technology.
\end{abstract}

\maketitle

\section{Introduction}

A highly accurate physical quantity estimation is of great importance and has pushed forward the development of science and technology. In classical physics, the   estimation precision is bounded by the standard quantum limit (also named as shot noise limit) with $\Delta x\sim1/\sqrt{N}$ scaling, where $\Delta x$ is the fluctuation of the estimated parameter $x$ and $N$ is the number of resource employed. By use of the quantum effects, the standard quantum limit can be promoted to the Heisenberg limit where the precision will achieve $\Delta x\sim1/N$. The quantum metrology has been widely used in many fields, such as gravity wave detection~\cite{Caves,Aasi,Grote}, radar~\cite{Barzanjeh}, quantum sensing~\cite{Degen,Arrad}, optical imaging~\cite{Xiaoming,Fang,FFujimoto}, phase estimation~\cite{Humphreys,OH}, as well as atomic clock~\cite{Borregaard,Kruse}.

Entangled states are usually utilized to improve the parameter estimation accuracy
and attain the fundamental Heisenberg scaling allowed by quantum mechanics. However, we have to face two challenges. One challenge is the difficulty in preparing entangled states. For atom or artificial atom systems, only some few-body entangled states, such as Bell states, W states as well as Greenberger-Horne-Zeilinger states, have been successfully prepared in experiments~\cite{Dicarlo,molmer,Dicarlo1,Quiroga,Lin}. Motivated by the applications in quantum communication~\cite{Gisin}, people have made great efforts to prepare the eight and ten (or even more) photons entangled states~\cite{Walther,Zhao,Su,Pan1,Pan2}, but the photon number is still not {large} enough for performing quantum metrology. The other challenge is the unavoidable interaction between the system and the environment, which destroys the entanglement and therefore limits the estimation accuracy. To deal with this issue, dynamical decoupling~\cite{Degen,Yang}, feedback control~\cite{Zheng,Hirose,Liu} and many other approaches have been developed. Moreover, non-Markovian effect is also shown to be effective to maintain entanglement-induced high measurement precision~\cite{An1,An2}. Most of the above works focus on how to prepare or protect entangled states for quantum metrology.  In an alternative way, it is natural to ask how to perform a high precision parameter estimation which beats the Heisenberg limit without preparing entangled states~\cite{Higgins}.

To address such a problem, we propose a cavity QED scheme, where a series of two-level atoms randomly pass through a single-mode cavity~\cite{Liao}. By preparing the atom with some coherence initially, a recent experiment~\cite{Kim} has demonstrated the
single-atom super-radiance effect. That is, the steady state average photon number of the cavity is proportional to the square of (but not linearly dependent on) the number of the effective coupling atoms $N_c$ during the lifetime of a photon.  It motivates us to estimate the physical parameters (for example, the atom-cavity coupling strength, which is proportional to the atomic dipole moment) through measuring the photon number of the cavity field. Our results show that, the quantum metrology with the assistance of super-radiance~\cite{Paulisch,Wang} will achieve the Heisenberg limit. Here, we have regarded the atoms instead of the photons as the prepared source, but the final measurement is performed on the photons in the cavity. {So, the Heisenberg limit here means that the root-mean-square fluctuation of the atom-cavity coupling strength is proportional to $1/N_c^2$. The advantages of our scheme compared with other proposals are: (I) We need not to prepare the atomic entangled states initially before they enter the cavity. (II) Since we measure the photon number of the cavity field in the steady state, it is also not necessary to maintain the atomic entanglement, which is generated by their coupling to the cavity field.}

{In our paper, we firstly obtain the average values of the operators of the cavity field in its steady state by solving the effective master equation. Then, we discuss the dependence of the root-mean-square of the fluctuation of the atom-cavity coupling strength on $N_c$. Furthermore, we reconstruct the density matrix of the steady state for the cavity with the assistance of the Gaussion state theory~\cite{gau,gau1}. }  We find that the initial atomic coherence, which induces an effective driving to the cavity mode, serves as an core factor in our high-precision quantum metrology scheme. When the coherence is absent, we show that the steady state of the cavity is a thermal state, with the equilibrium temperature closed to zero and the precision of the parameter estimation will be bounded by the standard quantum limit. On the contrary, when the atomic coherence is present, the steady state {of the cavity} becomes a displaced thermal state (the details will be shown below) with a large amount of excited photons which is proportional to $N_c^2$. More interestingly, the steady state {of the cavity} can be described as an incoherent superposition of orthogonal states, and the superposition probabilities are dependent on the estimated parameter. {Meanwhile, the major component of the steady state is a coherent state with the average photon number proportional to $N_c^2$, and it makes an irreplaceable contribution to the Heisenberg limit in quantum metrology.}

The rest of the paper is organized as follows. In Sec.~\ref{model}, we present our model and derive the master equation. In Sec.~\ref{limit}, we show that the initial atomic coherence will induce an effective driving to the cavity, which leads to the Heisenberg limit in the quantum metrology process. In Sec.~\ref{discussion}, we discuss the underlying physics behind the Heisenberg limit. In Sec.~\ref{conclusion}, we give a short summary. In appendix {A and B},  we present some detailed calculations.
\section{Model and master equation}
\label{model}
\begin{figure}[tbp]
\centering
\includegraphics[width=6cm]{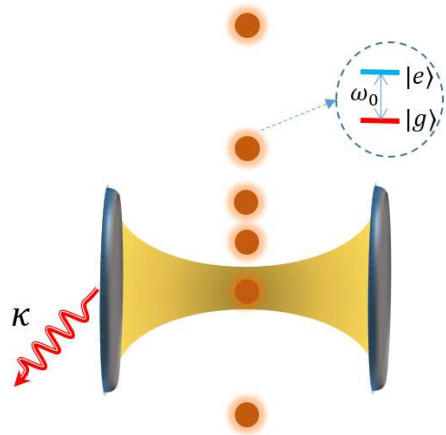}
\caption{Schematic diagram of our entanglement-free quantum metrology
model. A series of two-level atoms which are prepared in the same initial state randomly pass through a single-mode cavity one by one.}
\label{fly}
\end{figure}

We consider a cavity QED setup as shown in Fig.~\ref{fly}, which contains a single-mode cavity field of frequency $\omega$ and a series of identical two-level atoms whose energy separation between the excited states $|e\rangle$ and the ground states $|g\rangle$ are $\omega_{0}$. As shown in the figure, the two-level atoms are rapidly injected into the cavity with random time intervals to interact with the electromagnetic field in the cavity. We assume that the cavity mode is coupled to each atom within the same time duration  $\tau$, and there is at most one atom inside the cavity at any moment. In this paper, we will consider a simple situation where the two-level atoms are resonant with the single-mode cavity, that is, $\omega=\omega_{0}$. Then, in the interaction representation, the coherent coupling between a single two-level atom and the single-mode cavity field can be described  by the Jaynes-Cummings Hamiltonian (here and after, we set $\hbar=1$)
 \begin{equation}
V_{I}=g(\hat{a}\sigma_{+}+\hat{a}^{\dag}\sigma_{-}),
\label{1}
\end{equation}
and the evolution operator during the time interval $\tau$ is readily given by~\cite{Scully}
{\begin{eqnarray}
U(\tau)&=&\cos(g\tau\sqrt{\hat{a}\hat{a}^{\dagger}})|e\rangle\langle e|+
\cos(g\tau\sqrt{\hat{a}^{\dagger}\hat{a}})|g\rangle\langle g|\nonumber \\&&-i\frac{\sin(g\tau\sqrt{\hat{a}\hat{a}^{\dagger}})}
{\sqrt{\hat{a}\hat{a}^{\dagger}}}\hat{a}|e\rangle\langle g|-
i\hat{a}^{\dagger}\frac{\sin(g\tau\sqrt{\hat{a}\hat{a}^{\dagger}})}
{\sqrt{\hat{a}\hat{a}^{\dagger}}}|g\rangle\langle e|.\nonumber\\
\label{evolution}
\end{eqnarray}}
Herein, $g$ is the coupling strength between the cavity field and the two-level atom. $\hat{a}$ and $\hat{a^{\dag}}$ are respectively the annihilation and creation operators of the cavity field and obey the commutation relation $[\hat{a},\hat{a^{\dag}}]=1$. The Pauli operators $\sigma_{+}$ and $\sigma_{-}$ are defined as $\sigma_{+}=\sigma_{-}^{\dag}=|e\rangle\langle g|$.

We can denote the atomic injection rate as $r$, which represents the average number of atoms injected into the cavity per unit time interval. Then $r \delta t$ ($<1$ in our consideration) is the probability that an atom arrives at the cavity during the time interval $\delta t$. While $1-r \delta t$ is the probability that there is no atom in the cavity. In a realistic experimental scheme, the cavity {field} not only interacts with the injected atom, but also with the external environment. However, similar to the treatment in Refs.~\cite{Liao,Kim}, we neglect the effect of the environment when the atom is inside the cavity by assuming that the duration of the atom-cavity interaction is much shorter than that between two adjacent injections. Under such approximation, the time evolution of the density matrix of the cavity mode $\hat{\rho}(t)$ in a time interval $(t,t+\delta t)$ can be expressed as
\begin{equation}
\hat{\rho}(t+\delta t)=(1-r\delta t)[\hat{\rho}(t)+\mathcal{L}\hat{\rho}(t)\delta t]+r\delta t\mathcal{M}(\tau)\hat{\rho}(t).
\end{equation}
where
\begin{eqnarray}
\mathcal{M}(\tau)\hat{\rho}(t)&:=&{\rm Tr}_a[\hat{U}(\tau)\hat{\rho}(t)\otimes\hat{\rho}_a\hat{U}^{\dagger}(\tau)],\\
\mathcal{L}\hat{\rho}(t)&:=&\frac{\kappa}{2}[2\hat{a}\rho(t)\hat{a}^{\dagger}-
\hat{a}^{\dagger}\hat{a}\rho(t)-\rho(t)\hat{a}^{\dagger}\hat{a}].
\end{eqnarray}
Here, $\kappa$ is the decay rate of the cavity mode, $\hat{\rho}_a$ is the initial density matrix of the atom. ${\rm Tr}_a$ is the partial trace over the atom, and we have restricted the temperature to be zero. Neglecting the second order terms of $\delta t$ in the limit of $\delta t\rightarrow0$, we obtain the master equation:
\begin{eqnarray}
\dot{\hat\rho}&=&\lim_{\delta t\rightarrow0}\frac{\hat{\rho}(t+\delta t)-\hat{\rho}(t)}{\delta t}\nonumber\\&\approx&r[\mathcal{M}(\tau)-1]\hat{\rho}+\mathcal{L}\hat\rho.
\label{master}
\end{eqnarray}

\section{Coherence induced driving and Heisenberg limit}
\label{limit}
Similarly to the case in the recent coherent super-radiance  experiment~\cite{Kim}, we prepare all the atoms in the same initial state, which yields the initial density matrix (in the basis of $\{|e\rangle,|g\rangle\}$)
 \begin{equation}
\hat{\rho}_{a}=\left(\begin{array}{cc} p_{e} & \lambda \\\lambda^{*} & p_{g} \\\end{array}\right).
\label{3}
\end{equation}
Here, $p_e$ and $p_g$ are respectively the probability for the atom in its excited and ground states and $\lambda$ is the coherence of the two-level atoms. Without loss of generality, we will consider that $\lambda$ is real positive in the following of this paper.

In the presence of the atomic coherence ($\lambda\neq0$), the master equation (\ref{master}) can be further reduced by keeping up to the second order of $\tau$ to~\cite{Liao}
 \begin{equation}
\dot{\hat{\rho}}\approx i [\hat{\rho},H_{\rm{eff}}]+\mathcal {J}\hat{\rho},
\label{7}
\end{equation}
where the effective Hamiltonian is
 \begin{equation}
H_{\rm{eff}}=\xi \hat{a}^{\dag}+\xi^{*}\hat{a}\ \mathrm{with}\  \xi=rg\tau \lambda,
\label{8}
\end{equation}
and
 \begin{eqnarray}
\mathcal{J} \hat{\rho} &=& \frac{1}{2}\gamma_{1}(2\hat{a}^{\dag}\hat{\rho}\hat{a}
-\hat{a}\hat{a}^{\dag}\hat{\rho}-\hat{\rho}\hat{a}\hat{a}^{\dag})
\nonumber\\&+&\frac{1}{2} \gamma_{2}(2\hat{a}\hat{\rho}\hat{a}^{\dag}-\hat{a}^{\dag}\hat{a}\hat{\rho}
-\hat{\rho}\hat{a}^{\dag}\hat{a}),
\label{9}
\end{eqnarray}
is the modified dissipator with $\gamma_{1}=\alpha p_{e}, \gamma_{2}=\alpha p_{g}+\kappa$ and $\alpha=r(g\tau)^{2}$.

The Hamiltonian in Eq.~(\ref{8}) implies that the initial atomic coherence actually induces an effective coherent driving to the single-mode cavity field. This effective driving leads the steady state to deviate from the thermal state, whose equilibrium  temperature is closed to zero in our consideration (see the detailed analysis in Sec.~\ref{discussion}). Due to the effective driving, the cavity field will acquire appreciable excitations in the steady state. Since the intensity of the effective driving and hence the average photon number in the steady state is dependent on the atom-cavity coupling strength, this model provides us a path to measure or estimate the atom-cavity coupling strength.

{In the current scheme, the single-mode cavity can be regarded as a driven-dissipation
system. The dissipation originates from the external environments and the diagonal elements of atomic density matrix, and is described by the dissipator in Eq.~(\ref{9}). The driving comes from the off-diagonal elements of atomic density matrix (the atomic coherence), and is described by the Hamiltonian $H_{\rm eff}$ in Eq.~(\ref{8}).  In what follows, we will demonstrate that the effective driving plays a crucial role in achieving the Heisenberg limit in the quantum metrology.}

As shown in Appendix~\ref{A1}, under the steady state condition $\dot{\hat{\rho}}=0$, the average photon number is solved as
\begin{eqnarray}
&\langle \hat{a}^{\dag}\hat{a}\rangle =\frac{\gamma_{1}}{\gamma_{2}-\gamma_{1}}+\frac{4|\xi|^{2}}{(\gamma_{2}-\gamma_{1})^{2}}.
\label{12}
\end{eqnarray}
To demonstrate the effect of the injecting atoms on the steady state of the cavity, we now define the effective atom number $ N_c$ as
\begin{equation}
N_c:=\frac{r}{\kappa}.
\end{equation}
We note that $r$ is the atomic injection rate and $1/\kappa$ is the lifetime of the photon in the cavity, therefore $N_c$ is the effective atom number which can interact with the photon during its lifetime. In the parameter regime of
\begin{eqnarray}
N_{c} (g\tau)^{2}\ll 1,
\label{17}
\end{eqnarray}
the steady state photon number is approximated as
 \begin{eqnarray}
\langle \hat{a}^{\dag}\hat{a}\rangle & \approx &N_{c}(g\tau)^{2}p_{e}+4N_{c} ^{2} (g\tau)^{2}\lambda^{2}.
\label{18}
\end{eqnarray}

It is shown in the above equation that the steady average photon number is proportional to $g^2$, which implies that the coupling strength between the atom and the cavity mode can be detected by measuring the average photon number. According to the error transfer formula, the root-mean-square of the fluctuation $\Delta g^2$ associated with the photon number measurement can be expressed as~\cite{Dowling1,Dowling}

\begin{eqnarray}
\Delta g^{2} & = & \frac{\langle (\hat{a}^{\dag}\hat{a})^2 \rangle-\langle \hat{a}^{\dag}\hat{a}\rangle^{2}}{(\partial \langle \hat{a}^{\dag} \hat{a} \rangle / \partial g)^{2}}
\nonumber\\&\approx&\frac{1}{4\tau^{2} N_{c}[p_{e}+4N_{c}\lambda^{2}]},
\label{23}
\end{eqnarray}
where $\langle (\hat{a}^{\dag}\hat{a})^2\rangle$ can be obtained by solving the Langevin equation as shown in Appendix~\ref{A1}.

{In Fig.~\ref{vnc}, we plot the fluctuation $\Delta g^2$ as a function of $N_c$ in a log-log scale for different $\lambda$. When the atom is initially prepared without any coherence, that is, $\lambda=0$, we will obtain a standard quantum limit $\Delta g^2\sim 1/ N_c$. As for the nonzero initial atomic coherence ($\lambda\neq0$), the fluctuation behaves differently for small and large $N_c$. We first discuss the situation for large $N_c$, which satisfies $4N_c\lambda^2\gg p_e$. In this case, the first term in the denominator of Eq.~(\ref{23}) can be neglected safely, and it yields that $\Delta g^2\sim 1/ N_c^2$, which implies Heisenberg limit in quantum metrology. This can be observed clearly in Fig.~\ref{vnc}, where the curves can be approximated as straight lines for large $N_c$.  Furthermore, the slope of the line for $\lambda=0$ is about $-1$ while it becomes $-2$ for $\lambda\neq0$, implying a jump from the standard quantum limit to Heisenberg limit, with the assistance of the effective driving. For the case of small $N_c$ in which the first term in the denominator of Eq.~(\ref{23}) is comparable to the second term, the curves for $\lambda\neq0$ are a bit off the straight lines. It is meaningless to talk about the Heisenberg limit for such small $N_c$. However, the coherence induced driving still enhance the estimation accuracy dramatically. Taking $N_c=10$ as an example, the estimate precision is enhanced by about $9(23)$ times for $\lambda=0.3(0.5)$ compared with that for $\lambda=0$. Note that, in the recent experiment which demonstrates the single-particle superradiance~\cite{Kim}, the value of $N_c$ has been achieved by $7.3$. This means that the enhancement in quantum metrology by the atomic coherence can be observed experimentally, and will be more significant for large $N_c$, which yields the Heisenberg limit.}

\begin{figure}[tbp]
\centering
\includegraphics[width=8cm]{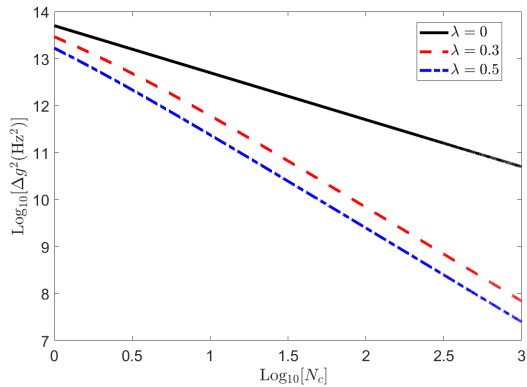}
\caption{The log-log plot of the root-mean-square of the fluctuation versus $N_c$. The parameters are set as {$\tau=100\rm{ns}$} and $p_{e}=0.5$.}
\label{vnc}
\end{figure}

\section{Discussion}
\label{discussion}
{As demonstrated above, the initial atomic coherence will effectively drive the cavity field and thus is beneficial for achieving the Heisenberg limit in quantum metrology. Our scheme differs from most of the traditional quantum precision measurement schemes in the following two aspects. Firstly, people usually prepared the entangled states for the employed source before parameterization (the parameterization is usually implemented through the dynamical evolution process), to achieve a higher parameter estimation accuracy, for example, the Heisenberg limit~\cite{Giovann,Humphreys,Dobrza}. In our scheme, the atoms only possess some coherence initially, but preparing the initial entangled states is not necessary. Secondly, in the traditional schemes, the states of the sources themselves (for example the atoms or photons in the interferometer) are measured after parameterization. In our scheme, we have considered the injected atoms as the source, and the final measurement is performed on the photons of the cavity field.  In such a situation, it is plausible to investigate the characterization of the steady state of the cavity and discuss the experimental feasibility.}

\subsection{Characterization of the steady state}

Remember that the dynamical behavior of the {cavity field is governed by the effective Hamiltonian} with a quadratic form [note that the dissipators in Eq.~(\ref{9}) can be obtained by regarding the cavity field to interact with the environments via a quadratic Hamiltonian], the steady state yields a Gaussian state. After some detailed calculations as shown in Appendix~\ref{A2}, the density matrix of the steady state is expressed as
\begin{equation}
\hat{\rho}=\hat{D}(\alpha_{0})\hat{\rho}_{T}\hat{D}^{\dagger}(\alpha_{0}),
\label{finalsteady}
\end{equation}
where
\begin{equation}
\alpha_0=-2iN_{c}\lambda g\tau,\label{alpha0}
\end{equation}
and
\begin{equation}
\hat{D}(\alpha_0)=\exp(\alpha_0\hat{a}^\dagger-\alpha_0^*\hat{a})
 \label{displace}
 \end{equation}
 is the displace operator. $\hat{\rho}_T$ is the thermal state
\begin{eqnarray}
\hat{\rho}_T&=&\frac{1+N_c(g\tau)^2(1-2p_e)}{1+N_c(g\tau)^2(1-p_e)}\nonumber\\&&\times\sum_{n=0}
\{[\frac{N_c(g\tau)^2p_e}{1+N_c(g\tau)^2(1-p_e)}]^n|n\rangle\langle n|\},
\end{eqnarray}
with $|n\rangle$ being the Fock state of the cavity field with $n$ photons. Similar to the previous discussion, we keep up to the first order of $N_c(g\tau)^2$, it yields
\begin{equation}
\hat{\rho}_T\approx[1-p_eN_c(g\tau)^2]|0\rangle\langle0|+p_eN_c(g\tau)^2|1\rangle\langle1|.
\label{thermal}
\end{equation}

When all of the atoms are prepared in the mixed state with $\lambda=0$, the
cavity is equivalently immersed in a thermal reservoir, and the effective driving
disappears, in that $\xi=0$ in Eq.~(\ref{8}). In this case, the steady state is the thermal equilibrium state, whose density matrix is expressed in Eq.~(\ref{thermal}).
It is noted that, the average photon number in the above thermal state is $\langle \hat{a}^\dagger \hat{a}\rangle=p_eN_c(g\tau)^2$, which is very small in our considered parameter regime. In other words, the cavity field will reach a thermal equilibrium state of nearly zero temperature when the atomic initial coherence is absent.

However, when the atoms possess some coherence initially, an effective driving field with intensity $\xi$ coexists with the reservoir. As a result, we find an extra displacement on the thermal state, the amplitude of the displacement $\alpha_0$ is proportional to the initial atomic coherence $\lambda$. Subsequently, the steady state will possess appreciable excitations. In the above discussions, we have named the state given by Eq.~(\ref{finalsteady}) as the ``displaced thermal state".

In a recent investigation of the single-atom super-radiance~\cite{Kim}, the authors kept up to the first order of $g\tau$, so that $\hat{\rho}_T\approx|0\rangle\langle 0|$, and the steady state was predicted to be the coherent state $\hat{\rho}\approx\hat{D}(\alpha_0)|0\rangle\langle 0|\hat{D}^{\dagger}(\alpha_0)=|\alpha_0\rangle\langle\alpha_0|$. However, in all of our previous calculations, we have always kept to the first order of $N_{c}(g\tau)^{2}$, it leads to the steady state
\begin{equation}
\hat{\rho}\approx[1-p_eN_{c}(g\tau)^{2}]|\alpha_{0}\rangle\langle\alpha_{0}|
+p_eN_{c}(g\tau)^{2}\hat{D}(\alpha_{0})|1\rangle\langle1|\hat{D}^{\dagger}(\alpha_{0}),
\end{equation}

It is clear that the steady state is an incoherent superposition of two orthogonal states  $|\psi_1\rangle=|\alpha_0\rangle$ and  $|\psi_2\rangle=\hat{D}(\alpha_0)|1\rangle$ with the superposition probabilities $p_1=1-x$ and $p_2=x$ respectively, where $x=p_eN_{c}(g\tau)^{2}\ll1$ in our consideration. Then, the average photon number in Eq.~(\ref{18}) is re-expressed as
\begin{equation}
\langle \hat{a}^\dagger \hat{a}\rangle=(1-x)\langle\psi_1| \hat{a}^\dagger \hat{a}|\psi_1\rangle+x\langle\psi_2| \hat{a}^\dagger \hat{a}|\psi_2\rangle,
\end{equation}
which is a weight summation of the average photon number in the two steady state components. Now, let us discuss the property of the fluctuation. The fluctuation for the state $|\psi_n\rangle$ is
\begin{eqnarray}
\Delta g^{2}_{n} & = & \frac{\langle \psi_{n}| \hat{a}^{\dag}\hat{a}\hat{a}^{\dag}\hat{a} |\psi_n\rangle-\langle \psi_n| \hat{a}^{\dag}\hat{a}|\psi_n\rangle^{2}}{(\partial \langle \psi_n| \hat{a}^{\dag} \hat{a} |\psi_n\rangle / \partial g)^{2}}
\nonumber\\&=&\frac{2n-1}{16N_{c}^{2}\tau^{2}\lambda^{2}},
\label{sep}
\end{eqnarray}
for $n=1,2$. {We emphasize that $\Delta g^{2}_{1}\approx \Delta g^{2}$ [$\Delta g^{2}$ is obtained in Eq.~(\ref{23})] in the condition of $N_c\gg1$. That is, the coherent state component in the steady state makes a dominant contribution to the Heisenberg limit in quantum metrology.}

Meanwhile, it is obvious that $\Delta g^{2}\neq(1-x)\Delta g^{2}_{1}+x\Delta g^{2}_{2}$. The reasons come from two aspects. One is the fact that  $\langle \hat{a}^\dagger\hat{a}\rangle^2\neq\langle\psi_1|\hat{a}^\dagger\hat{a}|\psi_1\rangle^2+ \langle\psi_2|\hat{a}^\dagger\hat{a}|\psi_2\rangle^2$. The more interesting reason comes from the dependence of $x$ on the estimated parameter $g$, which may play an important role in reaching the Heisenberg limit. To clarify this point, we just assume a quantum state given by the density matrix
\begin{equation}
\hat{\rho}'=(1-y)|\alpha_{0}\rangle\langle\alpha_{0}|
+y\hat{D}(\alpha_{0})|1\rangle\langle1|\hat{D}^{\dagger}(\alpha_{0}),
\end{equation}
which is an incoherent superposition state of $|\alpha_0\rangle$ and $\hat{D}(\alpha_0)|1\rangle$, with $y$ being independent of the estimated parameter $g$. Then, the fluctuation is
obtained as
\begin{equation}
\delta^{2}g^{\prime}=\frac{(2y+1)}{16N_{c}^{2}\tau^{2}\lambda^{2}}
+\frac{(y-y^{2})}{64N_{c}^{4}g^{2}\tau^{4}\lambda^{4}}.
\end{equation}
In the limit of large $N_c$, it will reach the Heisenberg limit when $y$ is also independent of $N_c$ and reach the standard quantum limit when $y$ is linearly dependent on $N_c$. Therefore, the dependence of the incoherence superposition probabilities for different components on the estimated parameter also plays an important role in a general quantum metrology process, and we will leave the more systematic investigations in the future work.

\subsection{Experimental feasibility}
At last, it is instructive to outline the working parameter regime in our scheme. From
Eq.~(\ref{23}), we note that the relative error satisfies $\Delta g/g\sim1/(N_cg\tau)$ when $N_c\gg1$ and $\lambda=1/2$. For the realistic experimental scheme, both of the two following conditions must be satisfied.

(1) A measurement process is only valid when the value of the fluctuation is much smaller than the measured value itself, that is, $\Delta g/g\ll1$, which leads to the condition
 \begin{equation}
 \frac{1}{N_c\tau}\ll g.\label{C1}
\end{equation}

(2) In our above discussions, we have imposed a strong limitation that there is at most one atom in the cavity at any moment, so that the time interval between two neighboring atom injections should be much longer than the atom-cavity interaction time,  that is, $1/r\equiv1/(N_c\kappa)\gg\tau$, which then yields
  \begin{equation}
 \kappa\ll\frac{1}{N_c\tau}.\label{C2}
\end{equation}
Combining the two conditions in Eqs.~(\ref{C1},\ref{C2}), it naturally requires $\kappa\ll g$, which is actually inside the strong coupling regime in the cavity-QED setup. Since the strong coupling in natural atom systems~\cite{Kimble0,Kimble,Kato} and the ultra-strong and deep-strong coupling in quantum circuit systems have both been realized~\cite{ultra6,ultra7}, we believe our high-precision measurement scheme based on coherence induced driving can be performed in the foreseeing experiments.

It should be noted that, in the recent single-atom super-radiance experiment~\cite{Kim}, the atom is prepared in the coherent superposition state $|\phi\rangle_a=\sin(\theta/2)|e\rangle+\cos(\theta/2)\exp(i\phi)|g\rangle$, where $\theta$ is the mixing angle and $\phi$ is the atomic phase imprinted by the pump laser. This phase is introduced to guarantee the sufficient interaction between the atom and the cavity field. In our theoretical studies, we have assume the phase to be zero, so that $p_e=\sin^2(\theta/2), p_g=\cos^2(\theta/2), \lambda=\sin(\theta)/2$. When the mixing angle is tuned to be $\theta=\pi/2$, the initial coherence achieves its maximum value, which will induce a strong effective driving to the cavity field and hence enhance the Heisenberg limit quantum metrology.

\section{Conclusion}
\label{conclusion}
In conclusion, we have demonstrated a quantum metrology scheme to beat the Heisenberg limit in a cavity-QED setup. Unlike previous schemes, {the entangled states of the employed atoms are not required initially before they enter the cavity}, and hence our scheme is simple and robust to the environment. In this scheme, the two-level atoms which serve as the source are randomly injected into the {leaky} single-mode cavity one by one and the steady state average photon number of the cavity is measured.  The effective coherent driving to the cavity field, which is induced by the initially atomic coherence, results in a displaced thermal state as the steady state. Benefiting from the large average photon number, which is proportional to $N_c^2$, in the steady state, we can perform a high-precision measurement on the atom-field coupling strength and the precision can achieve the Heisenberg limit.

{At last, we point out that the steady state of the atom-cavity system is actually an entangled state. On the one hand, the interaction between the atom and the cavity field will undoubtedly induce their entanglement. On the other hand, the cavity field as a data bus will also indirectly induce the entanglement between different atoms. However, the advantage of our scheme compared with those in Refs.~\cite{Paulisch,Wang} is that only the steady state of the cavity counterpart is measured, so the preparation (maintaining) of the atomic entanglement before (after) they interact with cavity field is not required. We hope the proposed scheme without entangled states preparation and protection based on the recent experiment~\cite{Kim} will stimulate further studies in quantum information processing and quantum metrology.}

\begin{acknowledgments}
 This work is supported by the National Natural Science Foundation of China (under Grant Nos. 11875011, 11705026, 11534002 and 11775048). The China Postdoctoral Science Foundation under Grant No. 2017M611293, The Educational Commission of Jilin Province of China under Grant No. JJKH20190266KJ and the Fundamental Research Funds for the Central Universities under Grant No. 2412017QD003.
\end{acknowledgments}

\appendix
\addcontentsline{toc}{section}{Appendices}\markboth{APPENDICES}{}
\begin{subappendices}
\section{Steady state average values}
\label{A1}
In Eq.~(\ref{7}), we have obtained the master equation of the system. Here, we give the derivation process of Eqs.~(\ref{23}) through the dynamical equations of the average values. With the formula $\langle \hat{O}\rangle={\rm Tr}(\hat{\rho}\hat{O})$, where $\hat{\rho}$ is the density matrix and $\hat{O}$ is an arbitrary operator, we will have
\begin{equation}
\dot{\bf{A}}=M{\bf A}+{\bf B}
\end{equation}
where
{\begin{eqnarray}
{\bf A}&=& [\langle (\hat{a}^{\dag}\hat{a})^2\rangle, \langle \hat{a}^{\dag}\hat{a}\hat{a}^{\dag} \rangle, \langle \hat{a}\hat{a}^{\dag}\hat{a} \rangle, \langle \hat{a}^{\dagger 2} \rangle,\nonumber\\&& \langle \hat{a}^{\dag}\hat{a} \rangle, \langle \hat{a}^2 \rangle, \langle \hat{a}^{\dag} \rangle,\langle \hat{a}\rangle]^T,\nonumber\\{\bf B}&=&(\gamma_{1},i\xi^{*}, -i\xi, 0, \gamma_{1}, 0,i\xi^{*},-i\xi)^T,
\end{eqnarray}}
and

\begin{equation}
M=\left(\begin{array}{cccccccc}
2\delta & -2i\xi & 2i\xi^{*} & 0 & s_1 & 0 & i\xi & -i\xi^{*}\\
0 & \frac{3\delta}{2} & 0 & -i\xi & 2i\xi^{*} & 0 &s_2& 0\\
0 & 0 & \frac{3\delta}{2} & 0 & -2i\xi & -i\xi^{*} & 0 & s_2\\
0 & 0 & 0 & \delta & 0 & 0 & 2i\xi^{*} & 0\\
0 & 0 & 0 & 0 & \delta & 0 & -i\xi & i\xi^{*}\\
0 & 0 & 0 & 0 & 0 & \delta & 0 & -2i\xi\\
0 & 0 & 0 & 0 & 0 & 0 & \frac{\delta}{2} & 0\\
0 & 0 & 0 & 0 & 0 & 0 & 0 & \frac{\delta}{2}
\end{array}\right)
\end{equation}
with $\delta=\gamma_1-\gamma_2, s_1=3\gamma_1+\gamma_2, s_2=\gamma_1+\gamma_2$.

The steady state solution of $M{\bf A}+{\bf B}=0$ gives the average values as
\begin{subequations}
\begin{eqnarray}
\langle \hat{a}\rangle&=&\frac{2i\xi}{\gamma_1-\gamma_2},\\
\langle \hat{a}^2\rangle&=&\frac{-4\xi^2}{(\gamma_1-\gamma_2)^2},\\
\langle \hat{a}^{\dag}\hat{a}\rangle &=&\frac{\gamma_{1}}{\gamma_{2}-\gamma_{1}}+\frac{4|\xi|^{2}}{(\gamma_{2}-\gamma_{1})^{2}},
\end{eqnarray}
\label{12de}
\end{subequations}
and
\begin{equation}
\langle (\hat{a}^\dagger \hat{a})^2\rangle=\frac{\gamma_1(\gamma_1+\gamma_2)}{(\gamma_1-\gamma_2)^2}-
\frac{4(3\gamma_1+\gamma_2)|\xi|^2}{(\gamma_1-\gamma_2)^3}
+\frac{16|\xi|^{4}}{(\gamma_{2}-\gamma_{1})^{4}}.
\end{equation}

Under the condition of $N_{c} (g\tau)^{2}\ll 1$, we will obtain Eq.~(\ref{23}).

\section{Gaussian steady state}
\label{A2}
In the above discussions, we have mentioned that the dynamics of the system is governed by a quadratic Hamiltonian, which means the single-mode cavity field will experience a Gaussian channel~\cite{gau}. Therefore, the steady state is undoubtedly a Gaussian state.  According to the results in Ref.~\cite{gau1}, the Gaussian state of a single-mode bosonic field (denoted by the annihilation and creation operators $\hat{a}$ and $\hat{a}^{\dagger}$) with frequency $\omega$  can be written as
\begin{equation}
\hat{\rho}=\hat{D}(z_{0})\hat{U}_0(r,\theta,\theta_{1})\hat{\rho}_{0}\hat{U}_0^{\dagger}(r,\theta,\theta_{1})
\hat{D}^{\dagger}(z_{0}),
\end{equation}
 where $\hat{\rho}_{0}=2\sinh(\beta_{T}/2)\exp[-\beta_{T}(\hat{a}^{\dagger}\hat{a}+1/2)]$
 is the thermal equilibrium state with the effective temperature  $\beta_{T}=\omega/k_{B}T$ (note that $\hbar$ has been set to be $1$), where $k_B$ is the Boltzmann constant. The operator $\hat{D}(z_0)$ is defined in Eq.~(\ref{displace}) and
 \begin{equation}
 \hat{U}_0(r_0,\theta_0,\theta_{1})=\exp[-\frac{r_0}{2}\exp(i\theta_0)\hat{a}
 ^{\dagger2}+h.c.]\exp(-i\theta_{1}a^{\dagger}a),
 \end{equation}
 with $r_0\geq 0, -\pi<(\theta_0,\theta_1)\leq \pi$. The values of
 $z_{0},r_0,\theta_0,\theta_{1}$  can be determined by the first
 and second order moments of the field operators $\hat{a}$ and $\hat{a}^\dagger$ as
\begin{equation}
\langle \hat{a}\rangle=z_{0},\,\langle \hat{a}^{2}\rangle=-2\mu_{A}^{*}+z_{0}^{2},\,\langle \hat{a}^{\dagger}\hat{a}\rangle=\tau_0-\frac{1}{2}+|z_{0}|^{2},
\end{equation}
and $\mu_{A}=\frac{Q}{4}\sinh(x_0)\exp(-i\theta_0),\,\tau_0=\frac{Q}{2}\cosh(x_0)$. The newly
introduced parameters are defined as $x_0:=2r_0,\, Q:=\coth(\beta_{T}/2)$.

Comparing with the steady state average values given by Eqs.~(\ref{12de}) in our system, we will obtain
\begin{equation}
r_0=0,\, Q=\frac{1+p_eN_{c}(g\tau)^{2}}{1+(1-2p_e)N_{c}(g\tau)^{2}},\,z_0=\alpha_0,
\end{equation}
and the values of $\theta_0$ and $\theta_1$ which do not affect the results can be taken as arbitrary real numbers. At last, we will obtain the steady state in Eq.~(\ref{finalsteady}).
\end{subappendices}

\end{document}